\documentclass[a4paper]{jpconf}
\usepackage{graphicx}
\begin{document}
\title{Future Reactor Experiments}

\author{Miao He}

\address{Institute of High Energy Physics, Beijing}

\ead{hem@ihep.ac.cn}

\begin{abstract}
The measurement of the neutrino mixing angle $\theta_{13}$ opens a gateway for the next generation experiments to measure the neutrino mass hierarchy and the leptonic CP-violating phase. Future reactor experiments will focus on mass hierarchy determination and the precision measurement of mixing parameters. Mass hierarchy can be determined from the disappearance of reactor electron antineutrinos based on the interference effect of two separated oscillation modes. Relative and absolute measurement techniques have been explored. A proposed experiment JUNO, with a 20 kton liquid scintillator detector of $3\%/$$\sqrt{E(MeV)}$ energy resolution, $\sim$~53 km far from reactors of $\sim$~36 GW total thermal power, can reach to a sensitivity of $\Delta\chi^{2}>16$ considering the spread of reactor cores and uncertainties of the detector response. Three of mixing parameters are expected to be measured to better than 1\% precision. There are multiple detector options for JUNO under investigation. The technical challenges are new type of PMTs with high efficiency and highly transparent liquid scintillator. Funding has been approved from Chinese Academy of Sciences. A similar proposal was from Korea known as RENO-50. Both of them are going to start data taking around 2020.
\end{abstract}

\section{Introduction}

The last unknown mixing angle $\theta_{13}$ has been measured by Daya Bay~\cite{dyb.prl}\cite{dyb.cpc} and other reactor~\cite{reno}\cite{dc} and accelerator~\cite{t2k}\cite{minos} experiments. It opens a new era of neutrino experiments, to determine mass hierarchy and search for charge-parity (CP) violation in the neutrino oscillation. The large value of $\theta_{13}$ makes both measurements easier. There are a couple of proposed long baseline accelerator experiments\cite{t2hk}~\cite{lbne}\cite{lbno} which have abilities to measure both mass hierarchy and CP phase. On the other hand, reactor experiments at a medium baseline can still play important roles, since the disappearance of reactor electron antineutrinos doesn't reply on the CP phase, which can give a clean measurement of mass hierarchy. Besides, current and future reactor experiments have abilities to improve the precision of four mixing parameters significantly, which can be used to validate the unitarity of the mixing matrix.

\section{Future sensitivity of ongoing reactor experiments}

It was pointed out the sensitivities of $\sin^{2}2\theta_{13}$ with 3 years data are 0.008, 0.03, 0.02 at 90\% C.L. for Daya Bay, Double Chooz and RENO, respectively~\cite{future.sens}. Daya Bay released the new result of electron antineutrino oscillation amplitude and frequency based on a spectral measurement at the NuFact 2013 conference\cite{dyb.shape1}~\cite{dyb.shape2}. The 1-$\sigma$ precision of $\sin^{2}2\theta_{13}$ has been improved to around 10\%, and the future precision can reach to be better than 4\%. Daya Bay also, for the first time, measured the neutrino mass splitting $\Delta m^{2}_{32}$ using reactor antineutrinos, and the result is consistent with the one from accelerator neutrinos~\cite{minos2}. RENO's current precision of $\sin^{2}2\theta_{13}$ is about 16\%~\cite{reno2} and it's expected to be improved to 7\% assuming the systematic uncertainty can reach to 0.005. The combined analysis of neutron capture on different targets in the Double Chooz far detector gives about 30\% precision on $\sin^{2}2\theta_{13}$ and it's going to be improved significantly with near detector running in 2014\cite{dc2}.

\section{Mass hierarchy by reactor neutrinos}

Mass hierarchy can be determined by the precision energy spectrum measurement of reactor neutrinos, and looking for the interference between two oscillation frequency components driven by $\Delta m^{2}_{31}$ and $\Delta m^{2}_{32}$, respectively~\cite{mh1}\cite{mh2}, where $\Delta m^{2}_{ij}$ represents the mass difference of two neutrino mass eigenstates $\Delta m^{2}_{i}$ and $\Delta m^{2}_{j}$. At a baseline of about 60 km, more than 20 oscillation cycles can be seen. In this case, a Fourier transform of the $L/E$ spectrum, where $L$ is the baseline and $E$ is neutrino energy, can enhance the information of oscillation frequencies and thus improve the sensitivity of mass hierarchy~\cite{zhanl.fourier1}\cite{fourier}. In the $\Delta m^{2}$ spectrum after the Fourier transform, the relative amplitude and position of the peak and valley give the discrimination power of mass hierarchy, without any precondition of the absolute measurement of $\Delta m^{2}_{31}$ or $\Delta m^{2}_{32}$~\cite{zhanl.fourier1}. On the other hand, adding the absolute value of $\Delta m^{2}$ can bring an additional sensitivity. A $\chi^2$ analysis shows that, with an improved measurement of $\Delta m^{2}_{\mu\mu}$ (which is an approximation of $\Delta m^{2}_{31}$ or $\Delta m^{2}_{32}$) by the accelerator experiments in the near future, we can achieve additional $\Delta\chi^2$ by $\sim$ 9 and 4 for the 1\% and 1.5\% relative errors of the $\Delta m^{2}_{\mu\mu}$ measurement\cite{liyf.chi2}.

The sensitivity is affected by the experimental conditions, such as the detector energy response and distribution of multiple reactor cores. Since the oscillation is very fast along with energy, the precision measurement of the energy spectrum requires $3\%/$$\sqrt{E(MeV)}$ energy resolution and no less than 50,000 detected antineutrino interactions~\cite{zhanl.fourier2}. In an ideal case of a single reactor core and a single detector, the baseline was optimized at about 52 km~\cite{liyf.chi2}. Multiple reactors cores with different baselines can reduce the oscillation structure. According to Ref.~\cite{liyf.chi2}, the baseline difference can't be more than 500 m. The energy nonlinearity is a challenge for the liquid scintillator detectors because it's difficult to determine the quenching effect and the Cherenkov contribution. However, since there are multiple oscillations which can be observed in the $L/E$ spectrum, and each of them carries the same information of $\Delta m^{2}$, this redundancy can be used to measure the energy scale at different energies, which is called self-calibration. Ref.~\cite{liyf.chi2} uses a quadratic function to fit the residual nonlinearity and shows that the sensitivity is almost not affected after the self-calibration.

\section{The Jiangmen Underground Neutrino Observatory}

The Jiangmen Underground Neutrino Observatory (JUNO) is a multipurpose neutrino-oscillation experiment designed to determine neutrino mass hierarchy and precisely measure oscillation parameters by detecting reactor antineutrinos from the Yangjiang and Taishan Nuclear Power Plants, observe supernova neutrinos, study the atmospheric, solar neutrinos and geo-neutrinos, and perform exotic searches, with a 20 kiloton (kton) liquid scintillator detector of unprecedented 3\% energy resolution (at 1 MeV) at 700-meter deep underground and has other rich scientific possibilities.

\begin{figure}[h]
\begin{center}
\includegraphics[width=25pc]{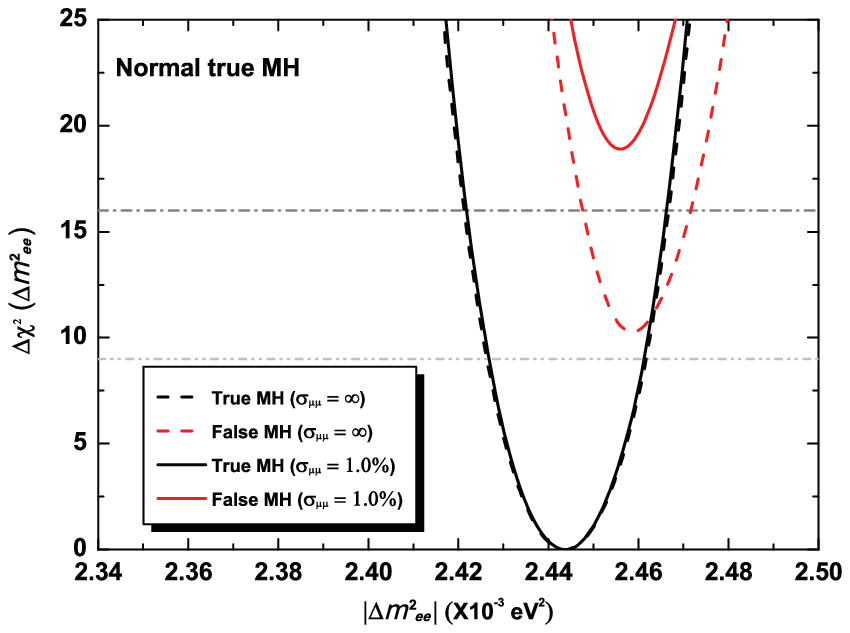}
\caption{\label{juno_mh}The mass hierarchy sensitivity for JUNO. Dash lines represent the relative measurement. Solid lines include the absolute value of $\Delta m^{2}_{\mu\mu}$ with a precision of 1\%.}
\end{center}
\end{figure}

The major goal of JUNO is determining neutrino mass hierarchy by precisely measuring the energy spectrum of reactor electron antineutrinos at a distance of ~53 km from the reactors. The relative measurement can reach to a sensitivity of $\Delta\chi^{2}>16$ in the idea case of a single reactor and a single detector, and $\Delta\chi^{2}>9$ considering the spread of reactor cores and uncertainties of the detector response. If the absolute value of $\Delta m^{2}_{\mu\mu}$ measured from accelerator experiments is included with a precision of 1\%, the sensitivity of mass hierarchy can be improved to $\Delta\chi^{2}>25$ and $\Delta\chi^{2}>16$ in the idea and real case, respectively, as shown in Figure~\ref{juno_mh}. JUNO is going to improve the precision of $\Delta m^{2}_{21}$, $\Delta m^{2}_{32}$ and $\sin^{2}\theta_{12}$ to be better than 1\%. Considering the precision of $\sin^{2}\theta_{13}$ can be measured to $\sim$4\% by Daya Bay, the unitarity of the neutrino mixing matrix can be probed to 1\% level. The expected precision of mixing parameters by JUNO is listed in Table~\ref{mix_par}.

\begin{table}[h]
\caption{\label{mix_par}Expected precision of mixing parameters by JUNO.}
\begin{center}
\begin{tabular}{lll}
\br
&Current&JUNO\\
\mr
$\Delta m^{2}_{21}$ & $\sim$3\% & $\sim$0.6\% \\
$\Delta m^{2}_{32}$ & $\sim$5\% & $\sim$0.6\% \\
$\sin^{2}\theta_{12}$ & $\sim$6\% & $\sim$0.7\% \\
$\sin^{2}\theta_{23}$ & $\sim$20\% & N/A \\
$\sin^{2}\theta_{13}$ & $\sim$4\% in a near future & $\sim$15\% \\
\br
\end{tabular}
\end{center}
\end{table}

\begin{figure}[h]
\begin{center}
\includegraphics[width=30pc]{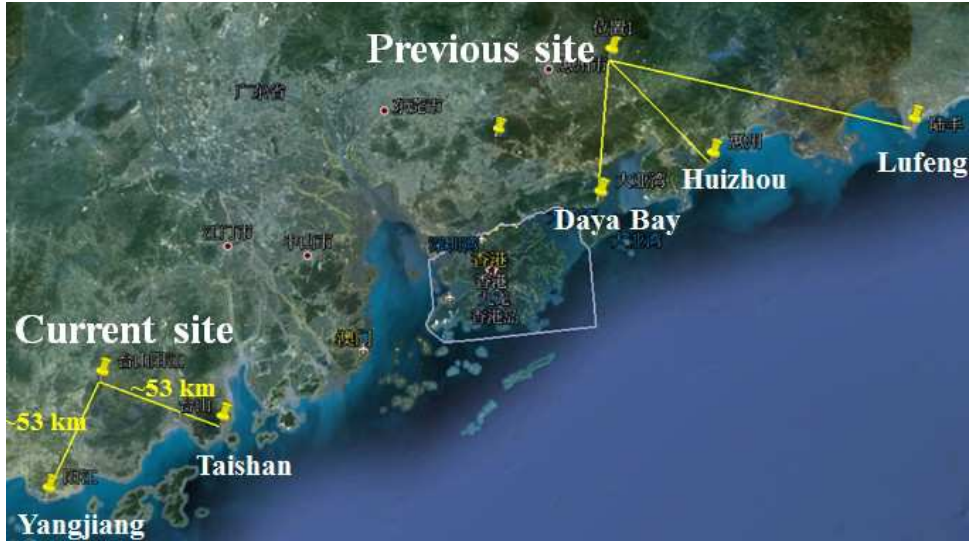}
\caption{\label{juno_site}Experiment site of JUNO. The previous site is near Daya Bay. That's the reason this experiment was known as Daya Bay II. The current site is in Jiangmen city.}
\end{center}
\end{figure}

JUNO is located in Kaiping, Jiangmen, in Southern China, as shown in Figure~\ref{juno_site}. It's about 53 km from the Yangjiang and Taishan nuclear power plants, both of which are under construction. The planned total thermal power of these reactors is 36 GW. There is no other nuclear power plant within 200 km. A 270 m high granite mountain provides good shielding of cosmic muons, which are the major sources of backgrounds. To further suppress muon induced backgrounds, the detector is designed to be located deep underground through a tunnel, and the total overburden will be 700 m rock. Experiment construction will start in 2013 and complete in 2019, including a tunnel, an underground experiment hall, a water pool, a central detector, a muon tracking detector, and some ancillary facilities.

\begin{figure}[h]
\begin{center}
\includegraphics[width=30pc]{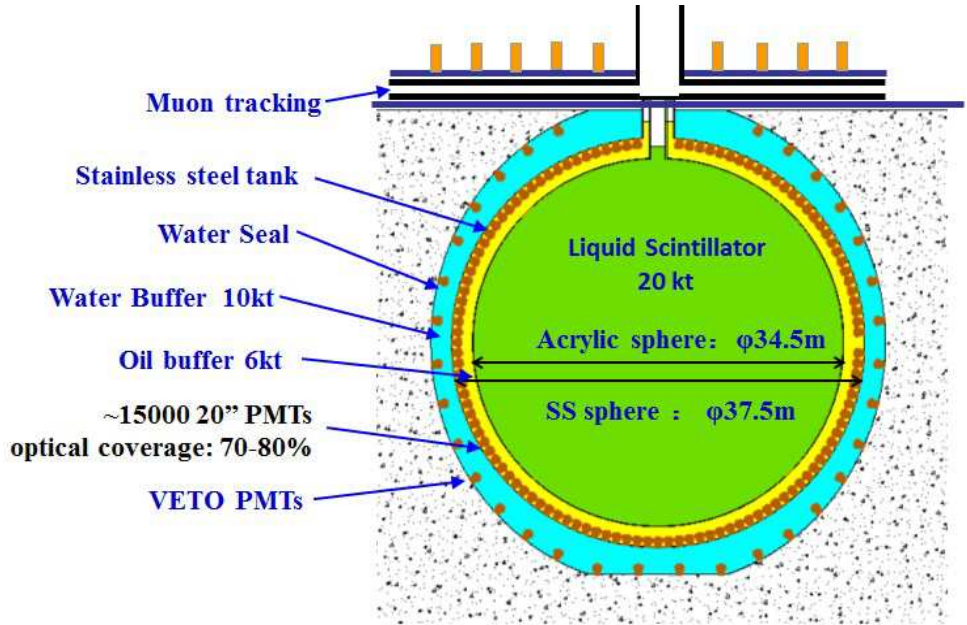}
\caption{\label{juno_det}A detector concept for JUNO.}
\end{center}
\end{figure}

The central detector concept includes two concentric spherical tanks located in a water pool, as shown in Figure~\ref{juno_det}. The inner acrylic tank is filled with 20 kton linear alkylbenzene (LAB) based liquid scintillator (LS). The outer stainless steel tank is filled with 6 kton mineral oil as buffer to protect LS from radioactivities. There are around 15,000 20" photomultiplier tubes (PMTs) installed in the internal surface of the steel tank. Since it's extremely difficult to build both large tanks at the same time, there are other options of the detector design. Option 1 removes the steel tank. The acrylic tank is directly put into water, which brings large pressure difference. Mineral oil is replaced by water in this option. PMTs can be installed in a steel frame in water. Option 2 removes the acrylic tank. Instead, small acrylic boxes filled with mineral oil can be made and installed as modules to contain single PMT or a group of PMTs. There are pipes at the back of each module for mineral oil filling and cabling. The leakage through cables is the major concern. Option 3 uses a balloon to replace the acrylic tank. Balloon is relatively cheap for construction and quick for installation. Experiences from Borexino and KamLAND are very encouraging. There are many technical details of film materials need to be considered, such as the transparency, radon permeability, and the leak check. Option 4 is actually a fall back plan for option 3. If the balloon is failed, the liquid scintillator has to be directly filled into the steel tank. In this case, PMTs are immersed in the liquid scintillator hence need special protection. On the other hand, since there is no mineral oil or water buffer, radioactivities from PMT will increase the trigger rate to more than 1 MHz, which can't be handled by the data acquisition system. An online data reduction algorithm was developed to reduce the trigger rate to less than 1 kHz based on the charge pattern detected by PMTs. However, the energy resolution is affected because of high probability of the overlap between antineutrino signals and radioactivities.

The water pool protects the central detector from natural radioactivity in surrounding rocks. It also serves as a water Cherenkov detector after being equipped with PMTs, to tag cosmic muons. There is another muon tracking detector on top of the water pool, used to improve muon detection efficiency and to get better muon tracking.

3\% energy resolution at 1 MeV corresponds to 1,200 photon electrons per MeV, which is a much better performance than the state of the art detector such as BOREXINO\cite{borexino} and KamLAND\cite{kamland}. The technical challenges are new type of PMTs with high efficiency and highly transparent liquid scintillator. The quantum efficiency of the photocathode made of super bialkali is expected to reach to 35\%. The traditional dynode can be replaced by the micro channel plate (MCP), which has near 4$\pi$ acceptance, receiving not only the transmission light but also the refection light with reflection photocathode at the bottom of PMT, hence largely improve the collection efficiency. A prototype of MCP-PMT has been made and under test. The ideal arrangement of 20" PMTs can reach to more than 80\% coverage. A mixture of 8" PMTs and 20" PMTs was considered, which can reach to the similar coverage as the ideal case, while the smaller PMT can provide better timing for event vertex reconstruction. An idea of adding reflection cones into the clearance was studied. With two thin acrylic panels with air gap, for uniformly distributed events, MC simulation shows $\sim$6\% increase on the total number of PE. Besides, reflecting to local PMTs won't impact on the vertex reconstruction. To reach a high transparency for the liquid scintillator, the production of LAB has been improved and the attenuation length of the raw liquid reached to 20.5 m at 430 nm wavelength. There are various of equipments produced for the purification, such as molecular distillation, vacuum distillation and filtration with the Al$_2$O$_3$ column. Currently, the attenuation length is increased to 24 m at 430 nm. Other methods such as lowering the temperature or optimizing the fluor concentration were studied to improve the light yield.

The reactor electron antineutrino interacts with the proton via the inverse $\beta$-decay (IBD) reaction in the liquid scintillator, and releases a positron and a neutron. The positron deposits its energy quickly, providing a prompt signal. The energy of positron carries most of the kinetic energy of the neutrino. The neutron is captured by a proton after an average time of 200 $\mu$s, then releases a 2.2 MeV gamma, providing a delayed signal. The coincidence of prompt-delayed signals provides a distinctive antineutrino signature. The estimated IBD reaction rate is ~40/day. The dominated background is accidental coincidence, coming from two uncorrelated background radiation interactions that randomly satisfy the energy and time correlation for the IBD antineutrino selection. It's designed to be less than 10\% of IBD signals and can be precisely measured in data. Other major backgrounds are introduced by cosmic muons, including cosmogenic $\beta$-n isotope $^9$Li/$^8$He and fast neutrons. Both of them are less than 1\% after appropriate muon veto.

JUNO got great support from Chinese Academy of Sciences with the Strategic Priority Research Program, which has been approved on February 1, 2013. The construction period is from 2013 to 2019. And the experiment will be in operation in 2020. There were two get-together meetings in January and July in 2013, and the next meeting will be in the experiment site Jiangmen the next January. Collaborators are welcome.

\section{RENO-50}

RENO-50 was proposed in South Korea\cite{reno50}. It plans an underground detector consisting of 18 kton ultra-low-radioactivity liquid scintillator and 15,000 20" PMTs, at 50 km away from the Hanbit nuclear power plant. The scientific goals include high precision measurement of $\theta_{12}$ and $\Delta m^{2}_{12}$, determination of mass hierarchy, study neutrinos from reactors, the Sun, the Earth, Supernova, and any possible stellar objects. The total budget is 100 million dollars for 6 years construction. Facility and detector construction will be from 2013 to 2018, and data taking will be started in 2019.

\section{Summary}

Ongoing reactor neutrino experiments in the world played important roles in measuring the smallest neutrino mixing angle $\theta_{13}$. They will continue running to improve the precision of $\sin^22\theta_{13}$ to be better than 4\%. Next generation reactor experiments will focus on mass hierarchy determination and precision measurement of mixing parameters. Science case is strong with significant technical challenges. JUNO was proposed a few years ago (known as Daya Bay II), now boosted by the large $\theta_{13}$. Funding has been approved from Chinese Academy of Sciences. A similar proposal was from Korea known as RENO-50. Both of them are going to start data taking around 2020.

\end{document}